\documentstyle[12pt]{article}

\begin{document}

\sloppy

\begin{flushright}{UT-699\\ Jan. '95}\end{flushright}

\vskip 1.5 truecm

\centerline{\large{\bf Comment on a mechanism of}}
\centerline{\large{\bf dynamical breaking of supersymmetry  }}
\vskip .75 truecm
\centerline{\bf Tomohiro Matsuda}
\vskip .4 truecm
\centerline {\it Department of Physics,University of Tokyo}
\centerline {\it Bunkyo-ku,Tokyo 113,Japan}
\vskip 1. truecm

\makeatletter
\@addtoreset{equation}{section}
\def\theequation{\thesection.\arabic{equation}}
\makeatother

\vskip 1. truecm

\begin{abstract}

We re-examine the so-called Nambu-Jona-Lasinio mechanism suggested
by Song, Xu and Chin in breaking the supersymmetry in the Wess-Zumino
model and show that this mechanism
 cannot be justified without assuming  special effects
between fermions.
The fermion condensation suggested by them corresponds to an unstable
vacuum configuration.
As a result, there is no fermion condensation and no supersymmetry
breaking in the model discussed by them.
\end{abstract}

\section{Introduction}
\hspace*{\parindent}
Recently it has been discussed in a series of papers by Song, Xu and Chin
\cite{xu} that spontaneous
supersymmetry breaking can be realized in a chiral
symmetric model without adding a Fayet-Iliopoulos or O'Raifeartaigh
term. In their analysis, the so-called Nambu-Jona-Lasinio mechanism
was used, and they suggest that the fermion pair condensation
induces a mass gap between supersymmetric partners.
If their mechanism really works, it would open many possibilities in
supersymmetric models.
The purpose of the present paper is to present the short comings of their
argument and clarify the physical backgrounds of it.
The main
point is very simple: They neglected the one-loop effects of bosonic
 particles.
Including these contributions correctly, we obtain the well-known
one-loop effective potential, and their solution corresponds to an
unstable configuration of this effective potential.

This paper is organized as follows. In section 2 we review the
construction of an effective potential in the Wess-Zumino model.
Then we re-examine the so-called
NJL method proposed in\cite{xu} and clarify the physical backgrounds.
 Concluding remarks are also given in section 3.

\section{Review of one-loop effective potential in WZ model}
The analysis of supersymmetry breaking in the Wess-Zumino model is as
old as the modern theory of supersymmetry\cite{zumino}.
Using a superfield method, Fujikawa and Lang\cite{fuji} constructed
a one-loop effective potential for the Wess-Zumino model and discussed
the stability of the supersymmetric vacuum.
Many authors, for example in\cite{miller}, later discussed this and
related topics.

For the notational convention, we use the two-component representation :
By explicitly separating the vacuum expectation values of bosonic fields,
we derive the one-loop effective potential by means of the tadpole
method\cite{tad} instead of the direct evaluation of it\cite{fuji}.

The starting Wess-Zumino Lagrangian for a chiral super multiplet is given by
\begin{eqnarray}
L&=&\Phi^{+}\Phi|_{\theta\theta\overline{\theta}\overline{\theta}}+\left[
\frac{1}{3!}\lambda\Phi^{3}|_{\theta\theta}+\frac{1}{2}
m\Phi^{2}|_{\theta\theta}+h.c.\right]\nonumber\\
&=&i\partial_{m}\overline{\psi}\overline{\sigma}^{m}\psi+\overline{A}
\Box{A}+\overline{F}F+\left[\frac{\lambda}{2}(A^{2}F-\psi\psi{A})+
m(AF-\frac{1}{2}\psi\psi)+h.c.
\right]
\end{eqnarray}
Shifting the bose fields of the theory in the fashion
\begin{eqnarray}
A&\rightarrow&A+a\nonumber\\
F&\rightarrow&F+f
\end{eqnarray}
we obtain
\begin{eqnarray}
L'&=&i\partial_{m}\overline{\psi}\overline{\sigma}^{m}\psi+\overline{A}
\Box{A}+\overline{F}F\nonumber\\
&&+\left[\eta(AF-\frac{1}{2}\psi\psi)+\frac{\lambda}{2}(AAF-\psi{\psi}A)
+\frac{\lambda}{2}fAA\right.\nonumber\\
&&\left.\hspace*{\fill}+F(ma+\frac{\lambda}{2}a^{2}-\overline{f})+A\eta{f}+
h.c.\right]
\end{eqnarray}
here we set
\begin{equation}
\eta=m+\lambda{a}.
\end{equation}
Before calculating the effective potential, we should derive the
propagators of the theory.
Extracting the quadratic part of the boson fields,
\vspace{1cm}
\begin{eqnarray}
S_{0}&=&\int{d}^{4}x\frac{1}{2}\Phi^{T}A\Phi+\Phi^{T}J\nonumber\\
&&\left\{
\begin{array}{l}
\Phi^{T}=(A,\overline{A},F,\overline{F})\\
J=(J,\overline{J},K,\overline{K})
\end{array}
\right.\nonumber\\
&&A=\left(
\begin{array}{llll}
-\lambda{f} & \Box & -\eta & 0 \\
\Box & -\lambda\overline{f} & 0 & \overline{\eta}\\
-\eta & 0 & 0 & 1 \\
0 & -\overline{\eta} & 1 0
\end{array}
\right)
\end{eqnarray}
the matrix $A$ is easily inverted to obtain
\begin{equation}
A^{-1}=\frac{1}{\Delta}\left(
\begin{array}{llll}
\lambda\overline{f} & \Box-\overline{\eta}\eta & \overline{\eta}(\Box
-\overline{\eta}\eta) & \lambda\overline{f}f \\
\Box-\overline{\eta}\eta & \lambda{f} & \lambda{f}\overline{\eta} &
\eta(\Box-\overline{\eta}\eta)\\
\overline{\eta}(\Box-\overline{\eta}\eta) & \lambda{f}\overline{\eta} &
\lambda{f}\overline{\eta}\overline{\eta} & -\lambda^{2}\overline{f}f
+ \Box(\Box-\overline{\eta}\eta)\\
\lambda\overline{f}\eta & \eta(\Box-\overline{\eta}\eta) & -\lambda^{2}
\overline{f}f+\Box(\Box-\overline{\eta}\eta) & \lambda\overline{f}
\eta\eta
\end{array}
\right)
\end{equation}
where
\begin{equation}
\Delta=(\Box-\overline{\eta}\eta)^{2}-\lambda^{2}\overline{f}f.
\end{equation}
The tree level generating functional is now given by
\begin{equation}
lnZ_{0}=-\frac{i}{2}\int{d}^{4}x(J^{T}A^{-1}J).
\end{equation}
Looking at $\frac{\delta^{2}ln(Z_{0})}{{\delta}J_{1}{\delta}J_{2}}|_{J=0}$
the propagators of the theory are obtained directly.

 Now let us derive the effective potential by means of the tadpole method.
According to ref.\cite{tad}, the following relation exists between the
derivative of effective potential and 1PI tadpole.
\begin{equation}
\frac{dV(\phi_{0})}{d\phi_{0}}=-\Gamma'^{(1)}
\end{equation}
In this expression
 $\phi_{0}$ means the vacuum expectation
value(vev) of the field $\phi$ which can be
any scalar field of the theory( in the present theory
 $\phi$ means $A$ or $F$, and
$\phi_{0}$ means $a$ or $f$), $\Gamma'^{(1)}$ is the 1PI tadpole that is
calculated after separating the vev and quantum fluctuation of the scalar
 fields as $\phi\rightarrow\phi+\phi_{0}$. So we use (2.3) to calculate
$\Gamma'^{(1)}$.
Using these relations, we obtain
\begin{equation}
\frac{V_{0}}{df}=ma+\frac{\lambda}{2}a^{2}-\overline{f}
\end{equation}
and
\begin{equation}
\frac{dV_{1}}{df}=-\frac{1}{2}\lambda^{2}
\int\frac{d^{4}p}{(2\pi)^{4}}
\frac{\overline{f}}{(p^{2}+\overline{\eta}\eta)^{2}-\overline{f}f
\lambda^{2}}.
\end{equation}
After integration we get
\begin{equation}
V_{0}=\left(ma+\frac{1}{2}\lambda a^{2}\right)f
-\overline{f}f+P(\overline{f},a,\overline{a})
\end{equation}
and
\begin{equation}
V_{1}=\frac{1}{2}\int\frac{d^{4}p}{(2\pi)^{4}}ln[(p^{2}+\overline{\eta}
\eta)^{2}-\lambda^{2}\overline{f}f]+H(\overline{a},a)
\end{equation}
where $P(\overline{f},a,\overline{a})$ and
$H(\overline{a},a)$ are  integration constants.
 We can impose supersymmetric boundary condition
\begin{equation}
V_{0}|_{f=0}=0
\end{equation}
and
\begin{equation}
V_{1}|_{f=0}=0.
\end{equation}
Then we recover the effective potential, which is also directly calculated
in ref.\cite{fuji},
\begin{equation}
V_{0}=\left[(ma+\frac{1}{2}\lambda a^{2})f+h.c.\right]-\overline{f}f
\end{equation}
and
\begin{equation}
V_{1}=\frac{1}{2}\int\frac{d^{4}p}{(2\pi)^{4}}ln\left[1-\frac{\lambda^{2}
\overline{f}f}{(p^{2}+\overline{\eta}\eta)^{2}}\right].
\end{equation}
The vacuum stability of this potential is well analyzed in
ref.\cite{fuji,miller}.
Eq.(2.17) can be evaluated as
\begin{eqnarray}
V_{1}&=&\frac{\pi^{2}}{(2\pi)^{4}}\left\{-\frac{1}{2}\lambda^{2}
|f|^{2}\left(ln\Lambda^{2}+\frac{1}{2}\right)\right.\nonumber\\
&&\left.+\frac{1}{2}|f|^{2}ln|
\eta|^{2}+\frac{1}{2}|\eta|^{2}[(1-x^{2})ln(1-x)+(1-x)^{2}
ln(1+x)]\right\}\nonumber\\
&&-(Z-1)|f|^{2}
\end{eqnarray}
where we set $x=\frac{|{\lambda}f|}{|\eta|^{2}}$, and $\Lambda$
stands for the ultra-violet cut-off.

We also added the wave function renormalization factor $Z$(in
the last term in (2.18)) in order to
absorb the infinity contained in $log\Lambda^{2}$. In order to avoid
the infrared singularity, which can appear because we set $m=0$ in the next
section, we renormalize the wave function at
\begin{equation}
|f|=0 \hspace{1cm} and \hspace{1cm}  |\eta|=M
\end{equation}
where $M$ has the dimensions of a mass. The wave-function
renormalization factor is then fixed as
\begin{equation}
Z=1-\alpha\left(ln\frac{\Lambda}{M^{2}}-1\right).
\end{equation}
The total effective potential up to one-loop level is now given by
\begin{eqnarray}
V_{eff}&=&-|f|^{2}\left(1-{\alpha}ln\frac{|\eta|^{2}}{M^{2}\lambda^{2}}\right)
\nonumber\\
&&+\frac{\alpha|\eta|^{4}}{2\lambda^{2}}
[(1+x)^{2}ln(1+x)+(1-x)^{2}ln(1-x)-3x^{2}]
\nonumber\\
&&+\lambda[(a_{1}^{2}-a_{2}^{2})f_{1}+2a_{1}a_{2}f_{2}]+2m(a_{1}f_{1}
+a_{2}f_{2}).
\end{eqnarray}
Here we set
\begin{eqnarray}
&&\left\{
\begin{array}{c}
f=f_{1}+if_{2}\\
a=a_{1}-ia_{2}
\end{array}
\right.,\nonumber\\
&&\alpha=\frac{\pi^{2}\lambda^{2}}{2(2\pi)^{4}}\ , \
x=\frac{|\lambda{f}|}{|\eta|^{2}}.\nonumber
\end{eqnarray}
In order to discuss the vacuum stability, we parametrize
$f_{1}$ and $f_{2}$
by
\begin{equation}
tan\beta=\frac{f_{1}}{f_{2}}
\end{equation}
and evaluate $V_{eff}$ at $\frac{\partial{V}_{eff}}{\partial{\beta}}=0$
( This corresponds to the direction of the valley of the effective
potential). We then find
\begin{eqnarray}
V_{eff}&=&-\frac{|\eta|^{4}x^{2}}{\lambda^{2}}
\left(1-{\alpha}ln\frac{|\eta|^{2}}{M^{2}}\right)
+\frac{\alpha}{2}|\eta|^{4}[(1+x)^{2}ln(1+x)+(1-x)^{2}ln(1-x)-3x^{2}]
\nonumber\\
&&+\frac{x|a||\eta|^{2}}{\lambda}\sqrt{\lambda^{2}(a_{1}^{2}+a_{2}^{2})+
2m{\lambda}a_{1}+m^{2}}
\end{eqnarray}
To take account of the two possible signs of the square-root, we extend
the range of $x$ to $-\infty<x<+\infty$. This potential develops an
imaginary part for $|x|>1$ and this means that the solution
\begin{equation}
|f|\ne0 \hspace{1cm} and \hspace{1cm} |\eta|=0
\end{equation}
is dynamically unstable.
We can find the stationary value of this effective potential
in the region $|x|\leq1$ assuming that $\alpha$ is small.
 The effective potential can be written as
\begin{eqnarray}
V_{eff}&\cong&-\frac{|\eta|^{4}x^{2}}{\lambda^{2}}
\left(1-{\alpha}ln\frac{|\eta|^{2}}{M^{2}}\right)
\nonumber\\
&&+\frac{x|a||\eta|^{2}}{\lambda}\sqrt{\lambda^{2}(a_{1}^{2}+a_{2}^{2})+
2m{\lambda}a_{1}+m^{2}}.
\end{eqnarray}
Taking the minimum of the potential ($\partial{V_{eff}}/\partial{x}=0$),
we obtain
\begin{equation}
V_{eff}=\frac{|a|^{2}}{4}
\frac{\lambda^{2}(a_{1}^{2}+a_{2}^{2})+
2m{\lambda}a_{1}+m^{2}}{1-{\alpha}ln\frac{|\eta|^{2}}{M^{2}}}
\end{equation}
for
\begin{equation}
x=\frac{1}{2}\frac{\lambda|a|\sqrt{\lambda^{2}(a_{1}^{2}+a_{2}^{2})+
2m{\lambda}a_{1}+m^{2}}}{|\eta|^{2}\left(
1-\alpha{ln}\frac{|\eta|^{2}}{M^{2}}\right)}
\end{equation}
This potential has its minimum at
\begin{eqnarray}
a_{1}=0,\ &&  a_{2}=0,\ \  and \ \  f=0 \nonumber
\end{eqnarray}
or
\begin{eqnarray}
a_{1}=-\frac{m}{\lambda}, \ && a_{2}=0, \ \ and \ \ f=0.
\end{eqnarray}
In both solutions, $f$ is zero and supersymmetry is not broken.
The second solution gives non-zero vev of $a$ but $f$ still remains zero: Two
solutions (2.28) are actually two stable physically
equivalent solutions, since one can pass from one to the other
by a redefinition of the fields\cite{zumino}.
When we consider the massless Wess-Zumino model in the next section,
 the second solution becomes
$a_{1}=0, a_{2}=0$ so the vev of $a$ remains zero.
Detailed study of this phenomenon
from another point of view is given in ref.\cite{zumino}.

Let us examine the physical meanings of this solution.
 At the tree level, the equation of motion for auxiliary
field is
\begin{equation}
F=\frac{1}{2}\lambda\overline{A}^{2}.
\end{equation}
At the first glance, this equation seems to suggest that if the
tree level potential
develops a non-zero vacuum expectation value$<\overline{A}>$,
$<F>$ becomes non-zero and the supersymmetry of the theory can be broken
spontaneously. But this does not happen. Including higher order
quantum corrections,
supersymmetry-breaking vacuum ($<F>=\frac{1}{2}\lambda<\overline{A}>^{2}$
and$<\overline{A}>$ is non-zero)
 becomes unstable and the supersymmetric vacuum ($<F>=0$) remains stable.

Furthermore, there is no $\Lambda$ dependence in the effective potential
after renormalization of the wave-function.

To analyze the behavior of the effective potential at small $|\eta|$
reliably, the renormalization group improvement of the effective potential
has also been discussed in ref.\cite{fuji}.
The effective potential for the massless theory is
\begin{eqnarray}
V_{eff}&=&-\frac{|\eta|^{4}x^{2}}{\lambda^{2}}
\left(1-{\alpha}ln\frac{|\eta|^{2}}{M^{2}}\right)
+\frac{\alpha}{2}|\eta|^{4}[(1+x)^{2}ln(1+x)+(1-x)^{2}ln(1-x)-3x^{2}]
\nonumber\\
&&+x|a|^{2}|\eta|^{2}
\end{eqnarray}
The stationary value of this potential in this region $|x|<1$ is
 estimated to be
\begin{equation}
V_{eff}=\frac{\lambda^{2}|a|^{4}}{4\left(1-\alpha
ln\frac{|\eta|^{2}}{M^{2}}\right)}
\end{equation}
at
\begin{equation}
x=\frac{\lambda|a|^{2}}{|\eta|^{2}\left(1-\alpha
ln\frac{|\eta|^{2}}{M^{2}}\right)}.
\end{equation}
Renormalization group improvement of $V_{eff}$ suggests that
\begin{eqnarray}
V_{eff}&\simeq&\frac{1}{4}(\lambda(M)|a|^{3})^{\frac{4}{3}}
\lambda(|a|)^{\frac{2}{3}}
\nonumber\\
&\simeq&\frac{1}{4}(\lambda(M))^{2}|a|^{4}\frac{1}{\left(1-3\alpha
ln\frac{|\eta|^{2}}{M^{2}}\right)^{\frac{1}{3}}}
\end{eqnarray}
with the running coupling
\begin{equation}
\lambda(|a|)=\frac{\lambda(M)}{\left[1-3\alpha ln\frac{|a|^{2}}{M^{2}}
\right]^{\frac{1}{2}}}.
\end{equation}
Note that the combination $\lambda(M)|a|^{3}$ is renormalization
group invariant in this theory.

$V_{eff}$ in (2.33) has a minimum at $|a|=0$ for which
$\lambda(|a|)\rightarrow 0$ and the analysis of $V_{eff}$
is reliable.
For $|a|\rightarrow 0$, $x\rightarrow 0$ in (2.33)  and
thus $|f|\rightarrow0$ and no supersymmetry breaking.

This explicit analysis, which is useful to the discussion in the
next section, is of course consistent with the analysis on
the basis of Witten index\cite{witten}.

For the discussion of the next section, we summarize the results
restricting ourselves to the massless Wess-Zumino model.
First, there is no supersymmetry-breaking vacuum.
Second, the vev of scalar field $A$ remains zero.

\section{The meaning of NJL method in WZ model}
In this section we re-examine
the physical backgrounds of the NJL method
proposed in ref.\cite{xu}. For convenience, we first recapitulate
the basic procedure in ref.\cite{xu}.

The same lagrangian (2.1)
is used, but at the first stage we  eliminate the
auxiliary field $F$ using the equation of motion. The result is
(we here set $m=0$)
\begin{equation}
L=i\partial_{m}\overline{\psi}\overline{\sigma}^{m}\psi+A^{*}\Box{A}
-[\frac{\lambda}{2}\psi\psi{A}+h.c.]-\frac{1}{4}\lambda^{2}|A|^{4}.
\end{equation}
The equations of motion are given by
\begin{equation}
\left\{
\begin{array}{ccc}
\Box{A}+\frac{1}{2}\lambda^{2}A^{*}AA+\frac{1}{2}
\lambda\overline{\psi}\overline{\psi} & = & 0\\
\Box{A^{*}}+\frac{1}{2}\lambda^{2}A^{*}A^{*}A+\frac{1}{2}
\lambda\psi\psi & = & 0\\
\left[i\partial_{m}\overline{\sigma}^{m}-\lambda{A}\right]\psi & = & 0
\end{array}
\right..
\end{equation}
Taking the vacuum expectation value of the first equation in (3.2),
 one obtains
\begin{equation}
\Box<A>+\frac{1}{2}\lambda^{2}<AAA^{*}>=-\frac{1}{2}
\lambda<\overline{\psi}\overline{\psi}>.
\end{equation}
Expansion of $<A^{*}AA>$ and $<\overline{\psi}\overline{\psi}>$
to the one-loop level(i.e., to the order of $\hbar$)
is given by
\begin{equation}
\left\{
\begin{array}{lll}
<A^{*}AA>&=&<A^{*}><A><A>+<A^{*}>\left[\frac{\frac{}{}}{\frac{}{}}
AAloop\right]+<A>\left[\frac{\frac{}{}}{\frac{}{}}
AA^{*}loop\right]\\
<\overline{\psi}\overline{\psi}>&=&\left[\frac{\frac{}{}}{\frac{}{}}
\overline{\psi}\overline{\psi}loop\right].
\end{array}
\right.
\end{equation}
Here the results of the one-loop diagrams are symbolically represented.
Then eq.(3.3) becomes, to the one-loop order,
\begin{eqnarray}
0&=&\Box{a}+\frac{\lambda^{2}}{2}aaa^{*}\nonumber\\
&&+\frac{\lambda^{2}}{2}a
\left[\frac{\frac{}{}}{\frac{}{}}AA^{*}loop\right]+
\frac{\lambda^{2}}{2}a^{*}
\left[\frac{\frac{}{}}{\frac{}{}}AAloop\right]+
\frac{\lambda}{2}
\left[\frac{\frac{}{}}{\frac{}{}}\overline{\psi}
\overline{\psi}loop\right]
\end{eqnarray}
Neglecting the tadpoles of the bosonic fields and setting $\Box a=0$
in (3.5),
 we get the same answer as in
ref.\cite{xu};
\begin{equation}
\lambda{a}aa^{*}+Tr\left[\frac{\lambda}{2}\int\frac{d^{4}p}{(2\pi)^{4}}
\frac{1}{i\partial_{m}\overline{\sigma}^{m}-\lambda{a}^{*}}\right]=0.
\end{equation}
which leads to the fermion pair condensation and
a mass gap between the supersymmetric partners\cite{xu}.
In fact, the above equation (3.6) can be rewritten as
\begin{equation}
|a|^{2}=4\int\frac{d^{4}p}{(2\pi)^{4}}\frac{1}{p^{2}-\lambda^{2}|a|^{2}}.
\end{equation}
This equation looks like a well-known mass-gap equation.
The integration requires an
 ultra-violet cut-off, so the solution ($a$) of the
self-consistent equation(3.7) depends on the  ultra-violet cut-off parameter.
Shifting the fields in the lagrangian as $A\rightarrow{A}+a$,
with $a$ given by eq.(3.7), we obtain the masses
\begin{eqnarray}
m_{A}^{2}&=&\frac{\lambda^{2}}{2}|a|^{2}\nonumber\\
m_{\psi}&=&\lambda|a|.
\end{eqnarray}
The supersymmetric partners thus appears to acquire  different masses.
This is the mechanism noted in ref.\cite{xu}.

But we must not neglect bosonic tadpoles. As discussed in
the previous section, the neglect of bosonic tadpoles in (3.5)
is not consistent with the expansion in $\hbar$ and the resulting
effective potential corresponds to the expansion around an unstable
vacuum (i.e., $x=\frac{1}{2}$ in (2.23)).
The meaning of the equation (3.3) is now clear:
This equation means that the derivative of the effective potential
is set to zero at the minimum,
i.e.,
$\frac{\partial(V^{0}+V^{one-loop})}{\partial{a}^{*}}|_{vac}=0$.
One can easily obtain (3.3) by applying the tadpole method (2.9) to
the variable $a$, not to $f$. Substituting $A$ in (3.1) as
$A\rightarrow A+a$ and using the
tadpole method, one obtains
\begin{eqnarray}
\frac{d(V_{0}+V_{1})}{da^{*}}&=&\frac{\lambda^{2}}{2}aaa^{*}\nonumber\\
&&+\frac{1}{2}\lambda\left[\frac{\frac{}{}}{\frac{}{}}\overline{\psi}
\overline{\psi}loop\right]
+\frac{1}{2}\lambda^{2}a\left[\frac{\frac{}{}}{\frac{}{}}
AA^{*}loop\right]
+\frac{1}{2}\lambda^{2}a^{*}\left[\frac{\frac{}{}}{\frac{}{}}
AAloop\right]\nonumber\\
\end{eqnarray}
The evaluation and integration of (3.9)
 is slightly complicated in the present calculational scheme
but the result is the same as (2.30)(see ref.\cite{miller}).
Of course, there is no cutoff dependence in the final result
which explicitly remains in the analysis of ref.\cite{xu}, nor supersymmetry
breaking induced by fermion pair condensation in the full effective
potential resulting from (3.9). The stationary point of the
effective potential  correspond to the supersymmetry
preserving point of (2.30).

In conclusion, we have shown that the supersymmetry breaking solution
in ref.\cite{xu} is a direct consequence of the neglect of one-loop
bosonic effects in the loop expansion of the effective potential.
Since no dynamical mechanism why the one-loop fermion effects should
 be retained and why the one-loop boson effects should be neglected is given
in ref.\cite{xu}, we conclude that the so-called Nambu-Jona-Lasinio
mechanism suggested there is not justified in the conventional
 framework
of field theory without assuming some special attractive
force between fermions.

\section*{Acknowledgment}
We thank K.Fujikawa and A.Yamada for many helpful discussions.

\end{document}